\documentstyle[12pt]{article}
\begin{document}
\pagestyle{plain}
\title{Electron in the Ultrashort Laser Pulse}
\author{Miroslav Pardy\\
Institute of Plasma Physics ASCR\\
Prague Asterics Laser System, PALS\\
Za Slovankou 3, 182 21 Prague 8, Czech Republic\\
and\\
Department of Physical Electronics \\
Masaryk University \\
Kotl\'{a}\v{r}sk\'{a} 2, 611 37 Brno, Czech Republic\\
e-mail:pamir@physics.muni.cz}
\date{\today}
\maketitle
\vspace{50mm}

\begin{abstract}
After the clasical approach to acceleration of a charged particle
by $\delta$-form impulsive force,
we consider the corresponding quantum theory based on the Volkov
solution of the Dirac equation. We determine the modified Compton formula
for frequency  of photons
generated by the scattering of the $\delta$-form laser pulse on the
electron in a rest.
\end{abstract}

\newpage

\baselineskip 15 pt

\section{Introduction}

The problem of interaction an elementary particles with the laser field is,
at present time, one of the most prestigeous problem in the particle
physics. It is supposed that, in the future, the laser will
play the same role in particle physics as the linear or circle accelerators
working in today particle laboratories.

One of the problem is acceleration of charged particles by the laser.
The acceleration effectiveness of the linear or circle accelerators is
limitied not only by geometrical size of them
but also by the energy loss of accelerated particles
which is caused by bremsstrahlung during the
acceleration. The amount of radiation follows  from the Larmor formula
for emission of radiation by accelerated charged particle (Maier, 1999).

In case of laser acceleration the classical idea is that
acceleration is caused by laser light as the periodic electromagnetic field.
However, it is possible to show that periodic electromagnetic wave does not
accelerate electrons in classical and quantum theory, because the
electric and magnetic components of the light field are mutually
perpendicular and it means the motion caused by the classical
periodic electromagnetic field is not linear but periodic (Landau and
Lifshitz, 1962).
Similarly, the quantum motion of electron in such a
wave firstly described by Volkov (1935) must coincide in the classical limit
with the clasical solution.

The situation changes if we consider laser beam as a system of
photons and the interaction of electron with laser light
is via the one-photon Compton process

$$\gamma + e \rightarrow \gamma + e,\eqno(1a)$$
or, the multiphoton Compton process

$$n\gamma + e \rightarrow \gamma + e,\eqno(1b)$$
where $n$ is a natural number.

The equation (1b) is the symbolic expression of the two different physical
processes. One process is the nonlinear Compton effect in which several
photons are absorbed at a single point, but only single high-energy photon
is emitted. The second process is interaction where electron scatters twice
or more as it traverses the laser focus. In our article the attention is
devoted to the nonlinear Compton process.

It is evident that acceleration by laser can be adequately
described only by quantum electrodynamics.
Such viewpoint gives us the motivation to
investigate theoretically the effectiveness of acceleration of charged
particles by laser beam.
The acceleration of charged particles by laser beam has been studied by many
authors (Tajima, 1979; Katsouleas et al., 1983; Scully et al., 1991; Baranova
et al., 1994) .
Many designs for such devices has been proposed.
Some of these are not sufficiently developed to be readily intelligible,
others seem to be fallacious and others are unlikely to be relevant to
ultra high energies. Some designs were developed only
to observe pressure of laser light on microparticles in liquids and gas
(Askin, 1970; Askin, 1972).

It is necessary to say that acceleration by the Compton effect differs
from the effect of light pressure which was considerred in past for instance
by Russian physicists Lebedev and Nichols and Hull (1903) and which
was also verified experimentally by these scientists.
The measurement consisted in determination of force acting on the torsion
pendulum.
It was confirmed that the pressure
is very small. Only after invention of lasers the situation
changed because of the very strong intensity of the laser light which can cause the
great pressure of the laser ray on the surface of the condensed matter.

Here, we consider the interaction of an
electron with a laser pulse.
First, we consider the clasical approach to acceleration of a charged particle
by $\delta$-form impulsive force.
Then, we discusse  the corresponding quantum theory based on the Volkov
solution of the Dirac equation.
We determine the modified Compton formula for frequency  of photons
generated by the scattering of the $\delta$-form laser pulse on the
electron in a rest.

\section{Classical theory of interaction of particle with an  impulsive force}

We idealize the impulsive force by the dirac $\delta$-function.
Newton's second law for the interaction of a massive particle with mass
$m$ with an impulsive force $P\delta(t)$ is as follows

$$m\frac {d^{2}x}{dt^{2}} = P\delta(t),\eqno(2)$$
where $P$ is some constant.

Using the Laplace transform on the last equation, with

$$\int_{0}^{\infty}e^{-st}x(t)dt = X(s), \eqno(3)$$

$$\int_{0}^{\infty}e^{-st}{\ddot x}(t)dt = s^{2}X(s) - sx(0) - {\dot x}(0),
\eqno(4)$$

$$\int_{0}^{\infty}e^{-st}\delta(t)dt = 1, \eqno(5)$$
we obtain:

$$ms^{2}X(s) - msx(0) - m\dot x(0) = P.\eqno(6)$$

For a particle starting from rest with $\dot x(0) = 0, x(0) = 0$, we get

$$X(s) = \frac{P}{ms^{2}},\eqno(7)$$
and using the inverse Laplace transform, we obtain

$$x(t) = \frac{P}{m}t\eqno(8)$$
and

$$\dot x(t) = \frac{P}{m}.\eqno(9)$$

Let us remark, that if we express $\delta$-function  by the relation
$\delta(t) = \dot \eta(t)$, then from equation (2) follows
$\dot x(t) = P/m$, immediatelly. The physical meaning of the quantity $P$
can be deduced from equation $F = P\delta(t)$. After
$t$-itegration we have $\int Fdt = mv = P$, where $m$ is mass of a body
and v its final velocity (with v(0) = 0). It means that value of
$P$ can be determined
a posteriory and then this value can be used in more complex equations
than eq. (2). Of course it is necessary to suppose  that $\delta$-form of the
impulsive force is adequate approximation of the experimental situation.

In case of the harmonic oscillator with the damping force and under
influence of the general force, the Newton law is as follows:

$$m\frac {d^{2}x(t)}{dt^{2}}  + b{\dot x}(t)
+ kx(t) = F(t).\eqno(10)$$

After application of the Laplace transform  and with regard to the same
initial conditions as in the preceding situation,
$\dot x(0) = 0, x(0) = 0$, we get the following algebraic equation:

$$ms^{2}X(s) +bsX(s) + kX(s) = F(s),\eqno(11)$$
or,

$$X(s) = \frac{F(s)}{m\omega_{1}}
\frac {\omega_{1}}{(s + b/2m)^{2} + \omega_{1}^{2}} \eqno(12)$$
with $\omega_{1}^{2} = k/m - b^{2}/4m^{2}$.

Using inverse Laplace transform denoted by symbol ${\cal L}^{-1}$ applied
to multiplication of functions $f_{1}(s)f_{2}(s)$

$${\cal L}^{-1}(f_{1}(s)f_{2}(s)) = \int_{0}^{t} d\tau
F_{1}(t-\tau)F_{2}(\tau),
\eqno(13)$$
we obtain with $f_{1}(s) =  F(s)/m\omega_{1}, \quad f_{2}(s) = \omega_{1}/
((s + b/2m)^{2} + \omega_{1}^{2}), \quad F_{1}(t) = F(t)/m\omega_{1},\\
F_{2}(t) = \exp{(-bt/2m)}\sin\omega_{1}t$.

$$x(t) = \frac {1}{m\omega_{1}}\int_{0}^{t}F(t-\tau)e^{-\frac {b}{2m}\tau}
\sin(\omega_{1}\tau)d\tau.\eqno(14)$$

For impulsive force $F(t) = P\delta(t)$ we have from the last formula

$$x(t) = \frac {P}{m\omega_{1}}e^{-\frac {b}{2m}t}\sin\omega_{1}t.
\eqno(15)$$

\section{Classical interaction of a charged particle with a laser pulse}

If we consider the $\delta$-form electromagnetic pulse, then we can write

$$F_{\mu\nu} = a_{\mu\nu}\delta(\varphi).\eqno(16)$$
where $\varphi = kx = \omega t - {\bf k}{\bf x}$. In order to obtain the
electromagnetic impulsive force in this form, it is necessary
to define the four-potential in the following form:

$$A_{\mu} = a_{\mu}\eta(\varphi),\eqno(17)$$
where function $\eta$ is the Heaviside unit step  function defined by
the relation:

$$\eta(\varphi) = \left\{ \begin{array}{c} 0, \quad \varphi <  0  \\
1, \quad \varphi \geq 0
\end{array} \right. .
\eqno(18)$$

If we define the four-potential by the equation (17), then the
electromagnetic tensor with impulsive force is of the form:

$$F_{\mu\nu} = \partial_{\mu}A_{\nu} - \partial_{\nu}A_{\mu} =
(k_{\mu}a_{\nu} - k_{\nu}a_{\mu})\delta(\varphi) =
a_{\mu\nu}\delta(\varphi).\eqno(19)$$

To find motion of an lectron in $\delta$-form electromagnetic force, we
must solve immediately either the Lorentz equation, or,
to solve Lorentz equation
in general with four potential $A_{\mu} = a_{\mu}A(\varphi)$
and then to replace the four-potential by the eta function.
Following Meyer (1971) we apply his method and then replace $A_{\mu}(\varphi)$
by $a_{\mu}\eta(\varphi)$ in the final result.

The Lorentz equation with $A_{\mu} = a_{\mu}A(\varphi)$ reads:

$$\frac {dp_{\mu}}{d\tau} = \frac {e}{m}F_{\mu\nu}p^{\nu}
= \frac {e}{m}(k_{\mu}a\cdot p - a_{\mu}k\cdot p)A'(\varphi).
\eqno(20)$$
where the prime denotes derivation with regard to $\varphi$,
$\tau$ is proper time and $p_{\mu} = m(dx_{\mu}/d\tau)$.
After multiplication of the last equation by $k^{\mu}$, we get with regard to
the Lorentz condition $ 0 = \partial_{\mu}A^{\mu} =
a^{\mu}\partial_{\mu}A(\varphi) = k_{\mu}a^{\mu}A'$, or,
$k\cdot a = 0$ and $k^{2} = 0$, the following equation:

$$\frac {d(k\cdot p)}{d\tau} = 0\eqno(21)$$
and it means that $k\cdot p$ is a constant of the motion and it can be defined
by the initial conditions  for instance at time $\tau = 0$. If we put
$p_{\mu}(\tau = 0) = p_{\mu}^{0}$, then we can
write $k\cdot p = k\cdot p^{0}$. At this moment we have:

$$k\cdot p = \frac{mk\cdot dx}{d\tau} = m\frac {d \varphi}{d\tau},
\eqno(22)$$
or,

$$\frac{d\varphi}{d\tau} = \frac {k\cdot p^{0}}{m}.
\eqno(23)$$

So, using the last equation and relation $d/d\tau =
(d/\varphi)d\varphi/d\tau$, we can write equation (20) in the form

$$\frac {dp_{\mu}}{d\varphi} = \frac {e}{k\cdot p^{0}}
(k_{\mu}a\cdot p - a_{\mu}k\cdot p^{0})A'(\varphi)\eqno(24)$$
giving (after multiplication by $a^{\mu}$)

$$\frac {d(a\cdot p)}{d\varphi} = - ea^{2}A',\eqno(25)$$
or,

$$a\cdot p = a\cdot p^{0} - ea^{2}A.\eqno(26)$$

Substituting the last formula into (24), we get:

$$\frac {dp_{\mu}}{d\varphi} = -e\left(a_{\mu} -
\frac {k_{\mu}a\cdot p^{0}}{k\cdot
p^{0}}\right)\frac {dA}{d\varphi} -
\frac {e^{2}a^{2}}{2k\cdot p^{0}}\frac {d(A^{2})}{d\varphi}
k_{\mu}.
\eqno(27)$$

This equation  can be immediately integrated to
give the resulting momentum in the
form:

$$p_{\mu} = p_{\mu}^{0} - e\left(
A_{\mu} - \frac {A^{\nu}p_{\nu}^{0}k_{\mu}}{k\cdot p^{0}}\right)
- \frac {e^{2}A^{\nu}A_{\nu}k_{\mu}}{2k\cdot p^{0}}.\eqno(28)$$

Now, if we put into this formula the four-potential (17)
of the impulsive force,
then for $\varphi\ > 0$ when $\eta > 1$, we get:

$$p_{\mu} = p_{\mu}^{0} - e\left(
a_{\mu} - \frac {a^{\nu}p_{\nu}^{0}k_{\mu}}{k\cdot p^{0}}\right)
- \frac {e^{2}a^{\nu}a_{\nu}k_{\mu}}{2k\cdot p^{0}},
\eqno(29)$$

The last equation can be used to determination of the magnitude of $a_{\mu}$
similarly as it was done in discussion to the eq. (2). It can be evidently
expressed as the number of $k$-photons in electromagnetic momentum. For
$\varphi < 0$, it is $\eta = 0$ and therefore $p_{\mu} = p_{\mu}^{0}$

It is still necessary to say what is the practical realization of the
$\delta$-form potential. We know from the Fourier analysis
that the Dirac $\delta$-function can be expressed by integral
in the following form:

$$\delta(\varphi) = \frac {1}{\pi}\int_{0}^{\infty}\cos(s\varphi)ds.
\eqno(30)$$

So, the $\delta$-potential can be realized as the continual
superposition of the harmonic waves. In case it will be not possible
to realize  experimentally it,
we can approximate the integral formula by the summation
formula as follows:

$$\delta(\varphi) \approx \frac {1}{\pi}\sum_{0}^{\infty}\cos(s\varphi).
\eqno(31)$$

\section{Volkov solution of the Dirac equation with Heaviside four-potential}

We know that the four-potential is inbuilt in the Dirac equation and we
also know that if the potential is dependent on $\varphi$, then,
there is explicite solution of the Dirac equation which  was found by
Volkov (1935) and which is called Volkov solution. The quantum mechanical
problem is to find solution of the Dirac equation with the $\delta$-form
four-potential (17) and from this solution determine the quantum motion of
the charged particle under this potential. Let us first remember the
Volkov solution of the Dirac equation

$$(\gamma(p-eA) - m)\Psi = 0. \eqno(32)$$

Volkov (1935) found the explicit solution of this equation for four-potential
$A_{\mu} = A_{\mu}(\varphi)$, where $\varphi = kx$. His solution
is of the form (Berestetzkii et al., 1989):

$$\Psi_{p} = R \frac {u}{\sqrt{2p_{0}}}e^{iS}  =
\left[1 + \frac {e}{2(kp)}(\gamma k)(\gamma A)\right]
\frac {u}{\sqrt{2p_{0}}}e^{iS},
\eqno(33)$$
where $u$ is an electron bispinor of the corresponding Dirac equation

$$(\gamma p - m)u = 0.
\eqno(34)$$

The mathematical object $S$ is the classical Hamilton-Jacobi function,
which  was determined in the form:

$$S = -px - \int_{0}^{kx}\frac {e}{(kp)}\left[(pA) - \frac {e}{2}
A^{2}\right]d\varphi.
 \eqno(35)$$

If we write Volkov  wave function $\Psi_{p}$ in the form (33), then, for the
impulsive vector potential (17) we have:

$$S = -px - \left[e\frac {ap}{kp} - \frac {e^{2}}{2kp}a^{2}\right]\varphi,
\quad R = \left[1 + \frac {e}{2kp}(\gamma k)(\gamma a)\eta(\varphi)\right].
\eqno(36)$$

Our goal is to determine acceleration generated by the electromagnetic field
of the $\delta$-form which means that the four-potential $A_{\mu}$
is the Heaviside step function (18).
To achieve this goal, let us define current density
(Berestetzkii et al., 1989) as follows:

$$j^{\mu} = {\bar \Psi}_{p}\gamma^{\mu}\Psi_{p},
\eqno(37)$$
where $\bar\Psi$ is defined as the transposition of (33), or,

$$\bar\Psi_{p} = \frac {\bar u}{\sqrt{2p_{0}}}\left[1 +
\frac {e}{2(kp)}(\gamma A)(\gamma k)\right]
e^{-iS}.
\eqno(38)$$

After insertion of $\Psi_{p}$ and $\bar\Psi_{p}$
into the current density, we have with $A_{\mu} = a_{\mu}\eta(\varphi),
\eta^{2} = \eta$:

$$j^{\mu} = \frac {1}{p_{0}}\left\{p^{\mu} - ea^{\mu} +
k^{\mu}\left(\frac {e(pa)}{(kp)} - \frac {e^{2}a^{2}}{2(kp)}\right)
\right\}.
\eqno(39)$$
for $\eta > 0$, which is evidently related to eq. (28).

The so called kinetic momentum corresponding to $j^{\mu}$ is as follows
(Berestetzkii et al., 1989):

$$J^{\mu} = \Psi^{*}_{p}(p^{\mu} - eA^{\mu})\Psi_{p})
= {\bar \Psi}_{p}\gamma^{0}(p^{\mu} - eA^{\mu})\Psi_{p}) = $$

$$\left\{p^{\mu} - eA^{\mu} +
k^{\mu}\left(\frac {e(pA)}{(kp)} - \frac {e^{2}A^{2}}{2(kp)}\right)
\right\} + k^{\mu}\frac
{ie}{8(kp)p_{0}}F_{\alpha\beta}(u^{*}\sigma^{\alpha\beta}u),
\eqno(40)$$
where

$$\sigma^{\alpha\beta} = \frac {1}{2}(\gamma^{\alpha}\gamma^{\beta} -
\gamma^{\beta}\gamma^{\alpha}).\eqno(41)$$

Now, we express the four-potential by the step function.
In this case the kinetic momentum contains the tensor $F_{\mu\nu}$
involving $\delta$-function.
It means that there is a singularity at point $\varphi = 0$. This
singularity plays no role in the situation for
$\varphi > 0$ because in this case the $\delta$-function is zero. Then, the
kinetic momentum is the same as $j^{\mu}$.

\section{Emission of photons by an electron moving in the impulsive field}

We know that the matrix element $M$ corresponding to the emission of
photon by electron in the electromagnetic field is as follows
(Berestetzkii et al., 1989):

$$M = -ie^{2}\int d^{4}x \bar \Psi_{p'}(\gamma e'^{*})
\Psi_{p}\frac {e^{ik'x}}{\sqrt{2\omega'}},
\eqno(42)$$
where $\Psi_{p}$ is the wave function of an electron before interaction with
the laser pulse and $\Psi_{p'}$ is the wave function of electron after
emission of photon with components $k'^{\mu} = (\omega', {\bf k}')$.
The quantity $e'^{*}$ is the four polarization vector of emitted photon.

We write the matrix element in the more standard form:

$$M = g \int d^{4}x\bar\Psi_{p'}O\Psi_{p}\frac {e^{ik'x}}{\sqrt{2\omega'}},
\eqno(43)$$
where  $O = \gamma e'^{*}$, $g = -ie^{2}$ in case of the
electromagnetic interaction and

$$\bar \Psi_{p'} = \frac {\bar u}{\sqrt{2p'_{0}}}\bar R(p')e^{-iS(p')}.
\eqno(44)$$

Using the above definitions, we write the matrix element in the
form:

$$M = \frac{g}{\sqrt{2\omega'}} \frac {1}{\sqrt{2p'_{0}2p_{0}}}
\int d^{4}x{\bar u}(p')\bar R(p')OR(p)u(p)e^{-iS(p') + iS(p)} e^{ik'x}.
\eqno(45)$$

The quantity $\bar R(p')$ follows immediately from eq. (33), namely:

$$\bar R(p') =\overline
{\left[1 + \frac {e}{2kp'}(\gamma k)(\gamma a)\eta(\varphi)\right]}  =
{\left[1 + \frac {e}{2kp'}(\gamma a)(\gamma k)\eta(\varphi)\right]}.
\eqno(46)$$

Using

$$-iS(p') + iS(p) = i(p'-p) + i(\alpha' - \alpha)\varphi, \eqno(47)$$
where

$$\alpha = \left(e\frac {ap}{kp} - \frac {e^{2}}{2}\frac {a^{2}}{kp}\right),
\quad \alpha' = \left(e\frac {ap'}{kp'} - \frac {e^{2}}{2}
\frac {a^{2}}{kp'}\right),\eqno(48)$$
we get:

$$M = \frac{g}{\sqrt{2\omega'}} \frac {1}{\sqrt{2p'_{0}2p_{0}}}
\int d^{4}x\bar u(p')\bar R(p')OR(p)u(p)
e^{i(p' - p)x}e^{i(\alpha' - \alpha)\varphi}
e^{ik'x}.
\eqno(49)$$

With regard to the mathematical relation
$\eta^{2}(\varphi) = \eta(\varphi)$, we can put

$$\bar R(p')OR(p) = A + B\eta(\varphi).
\eqno(50)$$
where

$$A = \gamma e'^{*}\eqno(51)$$
and

$$B = \frac {e}{2(kp)}(\gamma e'^{*})(\gamma k)(\gamma a) +
\frac {e}{2(kp')}(\gamma a)(\gamma k)(\gamma e'^{*}) + $$

$$\frac {e^{2}}{4(kp)(kp')}(\gamma a)(\gamma k)(\gamma e'^{*})
(\gamma k)(\gamma a).
\eqno(52)$$

The total probability of the emission of photons during the interaction of the
laser pulse with electron is as follows:

$$W = \int \frac {1}{2}\sum_{spin. polar.}\frac {|M|^{2}}{VT}
\frac {d^{3}p' d^{3}k'}{(2\pi)^{6}}.\eqno(53)$$

It is evident that the total calculation is complex and involves
many algebraic operations with $\gamma$-matices and $\delta$-functions.
At this moment we rescricte the calculations to the most simple approximation
where we replace the term in brackets in eq. (33) by unit and so
we write instead  of eq. (33):

$$\Psi_{p}\sim \frac {u}{\sqrt {2p_{0}}}e^{iS},
\eqno(54)$$
which is usually used in simmilar form for the nonrelativistic calculations
as it is discussed by Kreinov et al. (1997). Then,
in this simplified situation ${\bar R} OR$ reduces to $A = \gamma e'^{*}$ and

$$M  = \frac{g}{\sqrt{2\omega'}} \frac {1}{\sqrt{2p'_{0}2p_{0}}}
\int dx^{4}\bar u (\gamma e'^{*})u
e^{i(p' - p)x}e^{i(\alpha' - \alpha)\varphi}
e^{ik'x} = $$

$$\frac{g}{\sqrt{2\omega'}} \frac {1}{\sqrt{2p'_{0}2p_{0}}}
\bar u (\gamma e'^{*})u \delta^{(4)}(lk + p - p' - k'),
\eqno(55)$$
where

$$l = \alpha-\alpha'. \eqno(56)$$

One important step is the
determination of $W$ in the determination of trace, because according to the
quantum electrodynamics of a spin states it is possible to show that
(Berestetzkii et al., 1989)

$$\frac {1}{2}\sum_{spin. polar.}|M|^{2} =
\frac {1}{2}{\rm Tr}\left\{(\gamma p' + m)A(\gamma p + m)
\gamma^{0}A^{+}\gamma^{0}\right\},
\eqno(57)$$

In order to determine trace $Tr$ of the combinations of $\gamma$-matrix,
it is suitable to know some relations. For instance:

$${\rm Tr}(a\gamma)(b\gamma) = 4ab, \quad
{\rm Tr}(a\gamma)(b\gamma)(c\gamma) = 0,\eqno(58)$$

$${\rm Tr}(a\gamma)(b\gamma)(c\gamma)(d\gamma) = 4\left[(ab)(cd) - (ac)(bd)
+ (ad)(bc)\right].
\eqno(59)$$

Then,

$${\rm Tr}\left[(\gamma p' + m)A(\gamma p + m)\bar A\right] =
S_{1} + S_{2} + S_{3} + S_{4};\quad \bar A =
\gamma^{0}A^{+}\gamma^{0},\eqno(60)$$
where (using relations (58) and (59) and
$\bar {\gamma^{\mu}} = \gamma^{\mu}$ with $e'e'^{*} = -1$)

$$S_{1} = {\rm Tr}[\gamma p'A\gamma p \bar A] =
4\left[(p'e'^{*})(pe') + (pp') + (p'e')(pe'^{*})
\right]\eqno(61)$$

$$S_{2} = {\rm Tr}[mA\gamma p \bar A] = 0\eqno(62)$$

$$S_{3} = {\rm Tr}[m\gamma p'A \bar A] = 0\eqno(63)$$

$$S_{4} = {\rm Tr}[m^{2}A\bar A] = 4m^{2}(e'e'^{*}) = -4m^{2}.\eqno(64)$$

At this moment we can write probability of the process $W$ in the form:

$$W = \int \frac {1}{2}\sum_{spin.\; polar.}\frac {|M|^{2}}{VT}
\frac {d^{3}p'd^{3}k'}{(2\pi)^{6}} = $$

$$\int \frac {d^{3}p'd^{3}k'}{(2\pi)^{6}} \frac {1}{2}
(S_{1} + S_{2} + S_{3} + S_{4})
\frac {1}{(2\pi)^{2}}\delta^{(4)}(lk + p -p' -k') = $$

$$\int \frac {d^{3}p'd^{3}k'}{(2\pi)^{6}}\frac {1}{2}
\delta^{(4)}(lk + p - p' -k')
4\left\{(p'e'^{*})(pe') + (pp') - m^{2} + (p'e')(pe'^{*})) \right\}.
\eqno(65)$$

The presence of the $\delta$-function in the last formula is expression of
the conservation law $lk + p  = k' + p'$, which we write in the form:

$$lk + p - k' = p'.\eqno(66)$$

If we introduce the angle $\Theta$ between ${\bf k}$ and ${\bf k}'$, then,
with  $|{\bf k}| = \omega$  and  $|{\bf k}'| = \omega'$, we get from
the squared equation (66) in the rest system of electron, where
$p = (m,0)$, the following equation:

$$l\frac {1}{\omega'} - \frac {1}{\omega} = \frac {l}{m}(1 - \cos\Theta);
\quad l = \alpha - \alpha', \eqno(67)$$
which is modification of the original equation for the Compton process

$$\frac {1}{\omega'} - \frac {1}{\omega} = \frac {1}{m}(1 - \cos\Theta).
\eqno(68)$$

We see that the substantial difference between single photon interaction
and $\delta$-pulse interaction is the factor $s = \alpha - \alpha'$.

We know that the last formula of the original Compton effect can be written
in the form suitable for the experimental verification, namely:

$$\Delta \lambda = 4\pi\frac{\hbar}{mc}\sin^{2}\frac {\Theta}{2},
\eqno(69)$$
which was used by Compton for the verification of the quantum
nature of light (Rohlf, 1994).

We can express equation (67) in new form. From equation $lk  + p = k' + p'$
we get after multiplication it by $k$ in the rest frame of electron:

$$kp' = \omega m - \omega\omega'(1-\cos \Theta).\eqno(70)$$

Then, $l$ in eq. (67) is given by the formula $(a \equiv (v,{\bf w}))$:

$$l = \frac {2evm - e^{2}a^{2}}{2\omega m} - \frac {2eap' -
e^{2}a^{2}}{2\omega[m - \omega'(1-\cos \Theta)]}.
\eqno(71)$$

The equation (67)can be experimentally verified by the
similar methods which was
used by Compton for the verification of his formula. However,
it seems that the interaction of the photonic pulse substantially differs
from the interaction of a single photon with electron.

The equation $lk + p = k' + p'$ is the symbolic
expression of the nonlinear Compton effect in which several
photons are absorbed at a single point, but only single high-energy photon
is emitted. The second process, where electron scatters twice
or more as it traverses the laser focus is not considerred here.
The nonlinear Compton process was experimentally confirmed for instance by
Bulla et al. (1996).

\section{Discussion}

We have presented, in this article, the classical derivation of law of
motion of a charged particle accelerated
by $\delta$-form mechanical impulsive force
in case of the free particle and for the damped harmonic oscillator.
Then, we found solution of the Lorentz equation for motion of a charged
particle accelerated
by the electromagnetic pulse. From the quantum theory based on the Volkov
solution of the Dirac equation,
we determined the current density and kinetic momentum of an electron
accelerated by the laser pulse. The total probability of
emmision of photons during the interaction of the laser pulse with
electron was also  derived. It involves
the relation between initial momenta of particle and photon and final ones.

The present article is continuation of the author
discussion on laser acceleration
(Pardy, 1998; Pardy, 2001), where the Compton model of laser acceleration
was proposed.

The $\delta$-form laser pulses are here considerred as an idealization
of the experimental situation in laser physics. Nevertheless, it was
demonstrated theoretically that at present time the zeptosecond and
subzeptosecond laser pulses of duration $10^{-21} - 10^{-22}$ s can be
realized by the petawat lasers (Kaplan and Shkolnikov, 2002). It means that the generation of
the ultrashort laser pulses is the keen interest in development
of laser physics.

New experiments can be realized and new measurements performed by means of
the laser pulses, giving new results and discoveries.
So, it is obvious that the interaction of particles with the laser pulses
can form, in the near future, the integral part of the laser and
particle physics in such laboratories as ESRF, CERN, DESY, SLAC and so on.

\vspace{10mm}

\noindent
{\bf REFERENCES}

\vspace{7mm}

\noindent
Ashkin, A. (1970). Acceleration and trapping of particles by radiation
pressure, Phys. Rev. Lett. {\bf 24}, No. 4,  156-159.\\[2mm]
Ashkin, A. (1970). The pressure of laser light, Scientific
American {\bf 226} (2), 63-71.\\[2mm]
Baranova,  N. B.,  Zel'dovich B. Ya. and  Scully, M. O. (1994).  Acceleration of\\
charged particles by laser beams,  JETP {\bf 78}, (3) 249-258.
\\[2mm]
Berestetzkii, V. B, Lifshitz E. M. and  Pitaevskii, L. P. (1989). \\
{\it Quantum electrodynamics}, (Moscow, Nauka). \\[2mm]
Bulla, C et al. (1996). Observation of nonlinear effects in Compton scattering,
Phys. Rev. Lett. {\bf 76}, No. 17,  3116 - 3119.\\[2mm]
Kaplan, A. E. and  Shkolnikov, P. L.  (2002). Lasetron: A proposed source of
powerful nuclear-time-scale electromagnetic bursts,
Phys. Rev. Lett. {\bf 88}, No. 7,  074801-1 -- 4. \\[2mm]
Katsouleas T. and  Dawson, J. M. (1983). Unlimited electron \\
acceleration in laser-driven plasma waves, Phys. Rev. Lett. {\bf 51}, No. 5,
392-395. \\[2mm]
Kreinov,  V. P.,  Reiss, H. R. and Smirnov, M. (1997). {\it Radiative processes in
atomic physics}, (John Wiley \& Sons, Inc.).\\[2mm]
Landau L. D. and  Lifshitz, E. M. (1962). {\it The Classical \\
Theory of Fields}, 2nd ed.~(Pergamon Press, Oxford).\\[2mm]
Maier, M. (1991). Synchrotron radiation, in: CAS-PROCEEDINGS,\\
ed. Turner, S. CERN 91-04, 8 May, 97-115.\\[2mm]
Meyer, J. W. (1971). Covariant classical motion of electron in a laser beam,
Phys. Rev. D {\bf 3}, No. 2, 621 - 622. \\[2mm]
Nichols, E. F. and  Hull, G. F. (1903). The pressure due to radiation,
Phys. Rev. {\bf 17},  26-50; ibid. 91-104. \\[2mm]
Pardy, M. (1998). The quantum field theory of laser acceleration,
Phys. Lett. A {\bf 243}, 223-228. \\[2mm]
Pardy, M. (2001). The quantum electrodynamics of laser acceleration,
Radiation Physics and Chemistry {\bf 61}, 391-394.\\[2mm]
Rohlf, J. W. (1994). {\it Moderm Physics from $\alpha$ to $Z^{0}$}
(John Wiley \& Sons, Inc. New York). \\[2mm]
Scully, M. O. and   Zubary, M. S. (1991). Simple laser accelerator: \\
Optics and particle dynamics,  Phys. Rev. A {\bf 44}, No. 4, 2656-2663.\\[2mm]
Tajima, T. and  Dawson, J. M. (1979). Laser electron accelerator,\\
Phys. Rev. Lett. {\bf 43}, No. 4, 267-270.\\[2mm]
Wolkow, D. M. (1935). $\ddot{\rm U}$ber eine Klasse von L$\ddot{\rm o}$sungen
der Diracschen Gleichung,
Z. Physik, {\bf 94}, 250 - 260.

\end{document}